\documentstyle[twoside,fleqn,espcrc2,epsfig]{article}

\newcommand{\eq}{\begin{equation}}
\newcommand{\en}{\end{equation}}
\newcommand{\eqa}{\begin{eqnarray}}
\newcommand{\ena}{\end{eqnarray}}

\newcommand{\AmS}{{\protect\the\textfont2
  A\kern-.1667em\lower.5ex\hbox{M}\kern-.125emS}}

\title{A nonperturbative determination of $C_A$} \author{$\mbox{S.~Collins}^a$ and C.~T.~H.~Davies\address{Dept. of Physics an d Astronomy,
    Glasgow University, Glasgow, G12 8QQ, Scotland}, UKQCD Collaboration}

\begin{document}

\begin{abstract}
We describe a non-perturbative determination of $c_A$ using
correlators containing the axial-vector and pseudoscalar currents at
zero and non-zero momentum.  We apply the method of Bhattacharya et al
to extract $c_A$ from the requirement that the ratio of appropriate
correlators for the PCAC relation becomes independent of time in the
excited state region. We find that the result depends strongly on the
order of the derivatives used in the PCAC relation. We also find that,
using the lowest order derivatives, we cannot get a consistent value
of $c_A$ between zero and non-zero momentum cases.  The $c_A$ values
that we obtain as we improve the derivatives are consistent and
decrease in magnitude heading towards the perturbative result.
\end{abstract}

\maketitle

\section{Introduction}
The use of Symanzik-improved lattice actions and matrix elements is
widespread. However, with each improvement term added the
corresponding coefficient must be determined to enable discretisation
effects to be reduced to the desired level. Considering the light
hadron spectrum and matrix elements, the relevant $O(a)$ improvement
coefficients are, for the most part, only known to $1$-loop in
perturbation theory. A nonperturbative determination of these
coefficients is desirable and may be simpler than performing higher
loop perturbative calculations.

\begin{figure}
\begin{center}
\epsfig{file=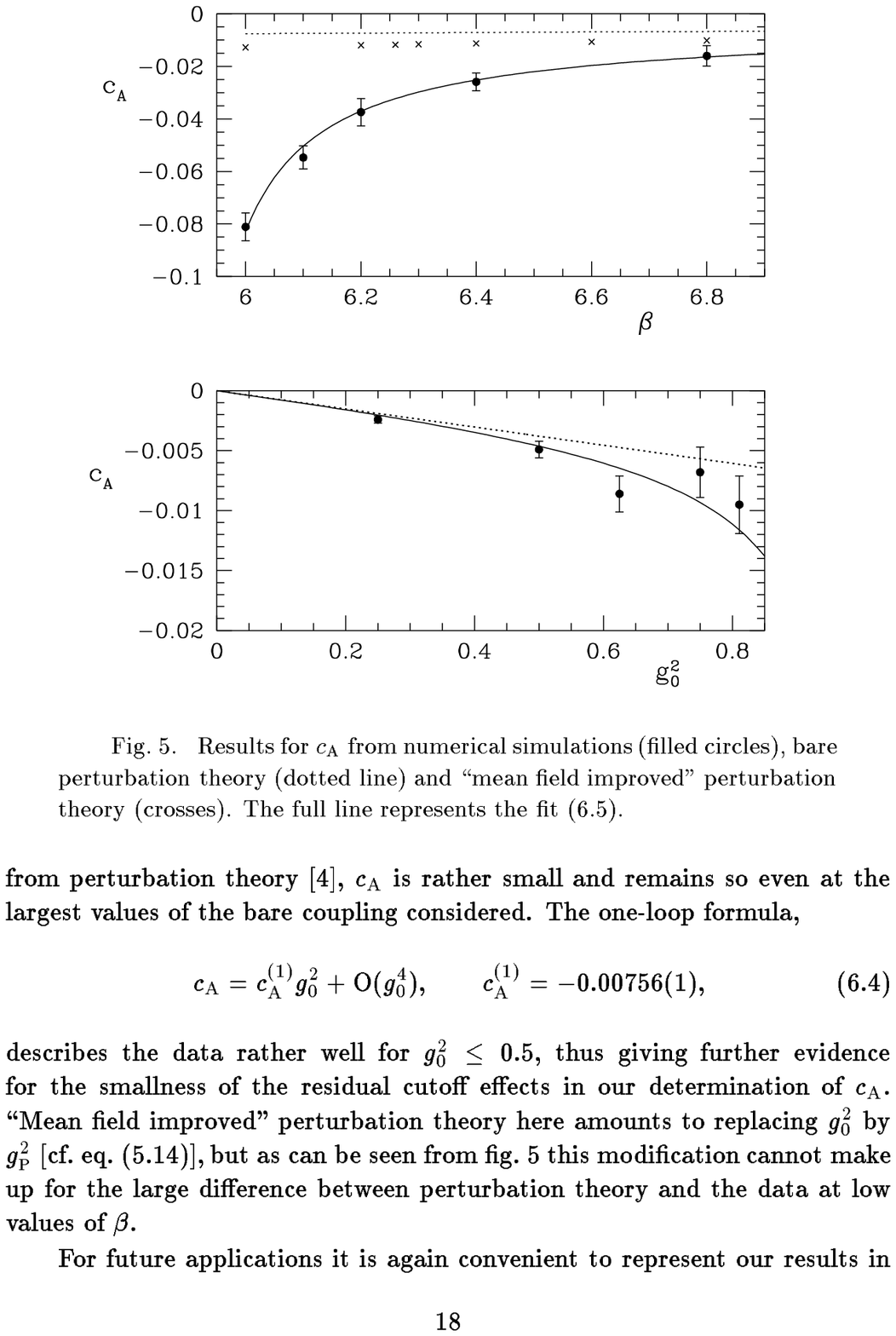,width=70mm,
clip= , bbllx=151pt,bblly=556pt,bburx=437pt,bbury=702pt} 
\vskip-1cm
\caption{$C_A$ extracted by the ALPHA collaboration. The circles indicate the
results of reference~\protect\cite{alpha}, the dotted line the results of 1-loop perturbation theory and the crosses the results from 1-loop
tadpole-improved perturbation theory.\label{falpha}}
\end{center}
\vskip-0.5cm
\end{figure}

The ALPHA collaboration, using Schr\"odinger functional techniques,
have calculated several $O(a)$ improvement coefficients
nonperturbatively. In most cases agreement is found with $1$-loop
(tadpole-improved) perturbation theory, or the discrepancy is
consistent with estimates of the size of omitted higher orders in
$\alpha$~(albeit in some cases assuming a slow convergence of the
perturbative series). However, for the $O(a)$ improvement of the
axial-vector current 
\begin{equation}
A_{\mu}\rightarrow A_{\mu}^I = A_{\mu} + c_A\partial_{\mu} P + O(a^2)\label{improv}
\end{equation}
where $A_\mu =
\bar{\psi}\gamma_\mu\gamma_5\psi$ and $P=\bar{\psi}\gamma_5\psi$, $c_A$ is found to be many times larger than the 1-loop perturbative
value even at reasonably high $\beta$, as shown in
figure~\ref{falpha}, taken from reference~\cite{alpha}.

$c_A$ appears in the expression for the pion decay constant, $f_\pi$,
\begin{equation}
f_\pi = (1+b_A)Z_A(f_\pi^{(0)} + c_A af_\pi^{(1)} +  O(a^2)
\end{equation}
and the bare quark mass~(from the PCAC relation)
\begin{equation}
2m = \frac{\partial_\mu<A_4 P> + c_Aa\partial^2_\mu<PP>}{2<PP>} +  O(a^2)
\end{equation}

Choosing the perturbative or nonperturbative value for $c_A$ leads to
significantly different values for these quantities at finite $\beta$.
Bhattacharya et. al. proposed an alternative method for determining
$c_A$, and other $O(a)$ coefficients in reference~\cite{lanl}. Here, we
apply their method, but with higher statistics than reported in
reference~\cite{lanl}. In addition, we extract $c_A$ at finite, as well as
zero, momentum.

\section{$C_A$ from the PCAC relation}
The PCAC relation in euclidean space
\begin{equation}
<\partial_{\mu} A_{\mu}(x) P> = 2m <P(x) P>\label{pcac}
\end{equation}
should hold for all $x$ on the lattice up to discretisation
terms. For simplicity we assign
\begin{eqnarray}
r(t) & = & \frac{<\partial_{\mu} A_{\mu}(t) P>}{<P(t)P>} \\
s(t) & = & \frac{<\partial_{\mu}^2 P(t) P>}{<P(t)P>}\label{es}
\end{eqnarray}
Thus, 
\begin{eqnarray}
2m & = & r(t) + \mbox{O(a)}\\
2m & = & r(t) + c_A s(t) + O(a^2). \label{im}
\end{eqnarray}
Equation~\ref{im} only holds if the $O(a)$-improvement term~(the
clover term) is included in the light quark action and the
coefficient, $c_{SW}$, is determined nonperturbatively.  The method of
Bhattacharya et. al. is to determine $c_A$ by minimizing the
dependence of $r(t)+c_As(t)$ on $t$ and, hence, reduce the
discretisation errors in this quantity.  Using this method, we
investigate the effect of improving $\partial_{\mu}$ on the
determination of $c_A$; discretisation errors in the lattice
derivatives may be the dominant source of lattice spacing errors in
the bare quark mass, and hence, the value obtained for $c_A$ may be
artificially high. 

Normally, the symmetric lattice derivatives
\begin{eqnarray}
\partial_{\mu} \rightarrow \Delta_{\mu}^{(+-)} & = & \frac{1}{2}(\delta_{\vec{x},\vec{x}+\hat{\mu}}-\delta_{\vec{x},\vec{x}-\hat{\mu}})\label{norm}\\
\partial^2_{\mu} \rightarrow \Delta^{(2)}_{\mu} & = & \delta_{\vec{x},\vec{x}+\hat{\mu}} - 2\delta_{\vec{x},\vec{x}} + \delta_{\vec{x},\vec{x}-\hat{\mu}}\nonumber
\end{eqnarray}
are used, which contain $O(a^2)$ errors. We will consider the effect of using
\begin{eqnarray}
\tilde{\Delta}_{\mu}^{(+-)} & = & \Delta_{\mu}^{(+-)} -\frac{1}{6}\Delta^+\Delta^{(+-)}\Delta^{-}\label{impd}\\
\tilde{\Delta}^{(2)}_{\mu} & =& \Delta_{\mu}^{(2)} -\frac{1}{12}\left[\Delta^+\Delta^-\right]^2\nonumber
\end{eqnarray}
which are correct up to $O(a^4)$. One can continue to correct to
$O(a^6)$, we denote these derivatives $\tilde{\Delta'}_{\mu}^{(+-)}$
and $\tilde{\Delta'}^{(2)}_{\mu}$.  As an additional constraint on
$c_A$, we investigate whether Lorentz invariance is restored to
sufficient accuracy i.e. that consistent values of $m$ are obtained at
zero and finite momentum.

\section{ Estimating $C_A$ using excited states}
The configurations and light quark propagators were provide by the
UKQCD collaboration.  The simulation parameters are given in
table~\ref{sim}. For more details see reference~\cite{ukqcd}. Note that the
$C_{SW}$ values correspond to those determined by the ALPHA
collaboration.

\begin{table}
\begin{center}
\begin{tabular}{ccccc}\hline
$\beta$ & Volume & $n_{confs}$ & $C_{SW}$ & $\kappa_l$  \\\hline
6.0 & $16^3\times 48$ & 496 & 1.77 & 0.13344 \\
6.2 & $24^3\times 48$ & 214 & 1.61 & 0.13460 \\\hline
$\beta$ & $\kappa_c$ & $\kappa_s(K)$ &  $a^{-1}(r_0)$\\\hline
6.0 & $0.135252(^{+16}_{-9})$ & $0.13401(^{+2}_{-2})$ & 2.12 & \\
6.2 & $0.135815(^{+17}_{-14})$ & $0.13495(^{+2}_{-2})$  & 2.91 & \\\hline
\end{tabular}
\caption{Simulation details.\label{sim}}
\end{center}
\vskip-0.5cm
\end{table}

\begin{figure}
\begin{center}
\epsfig{file=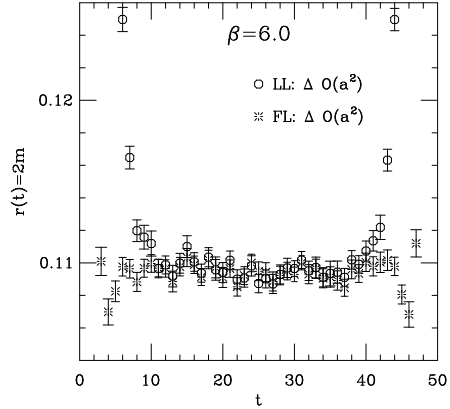,width=60mm}
\vskip-1cm
\caption{The ratio of correlators $r(t).$\label{figr}}
\end{center}
\vskip-0.5cm
\end{figure}

Figure~\ref{figr} shows the ratio of correlators, $r(t)$, for the data
set at $\beta=6.0$ for correlators local at the source and sink~(LL)
and those fuzzed at the source and local at the sink~(FL). The
$O(a^2)$ lattice derivatives, equation~\ref{norm}, were used. Towards
the center of the lattice $r(t)$ tends to a constant, $2m\sim
0.1096$. However, in the LL case, close to the origin there are
significant deviations from this value due to discretisation effects.
The much wider plateau in $r(t)$ for the FL correlators~(where the
fuzzing has been chosen to more or less eliminate excited state
contributions) show that the discretisation effects in $r(t)$ LL are
associated with excited states.  We can determine $c_A$ by adding the
term $c_A s(t)$ to reduce the discretisation effects at small $t$. To
do this we perform a {\it correlated} fit to $r(t)$ using $r(t) = 2m -
c_A s(t)$, to extract $2m$ and $c_A$. The values of $c_A$ obtained are
detailed in the next section.

\begin{figure}
\begin{center}
\epsfig{file=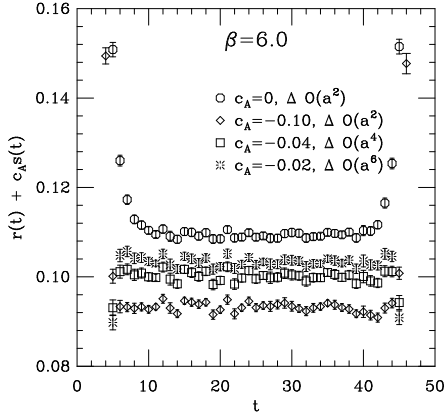,width=60mm}
\vskip-1cm
\caption{The $O(a)$-improved bare quark mass as determined using from
LL correlators.\label{cmass}}
\end{center}
\vskip-0.5cm
\end{figure}

\begin{figure}
\begin{center}
\epsfig{file=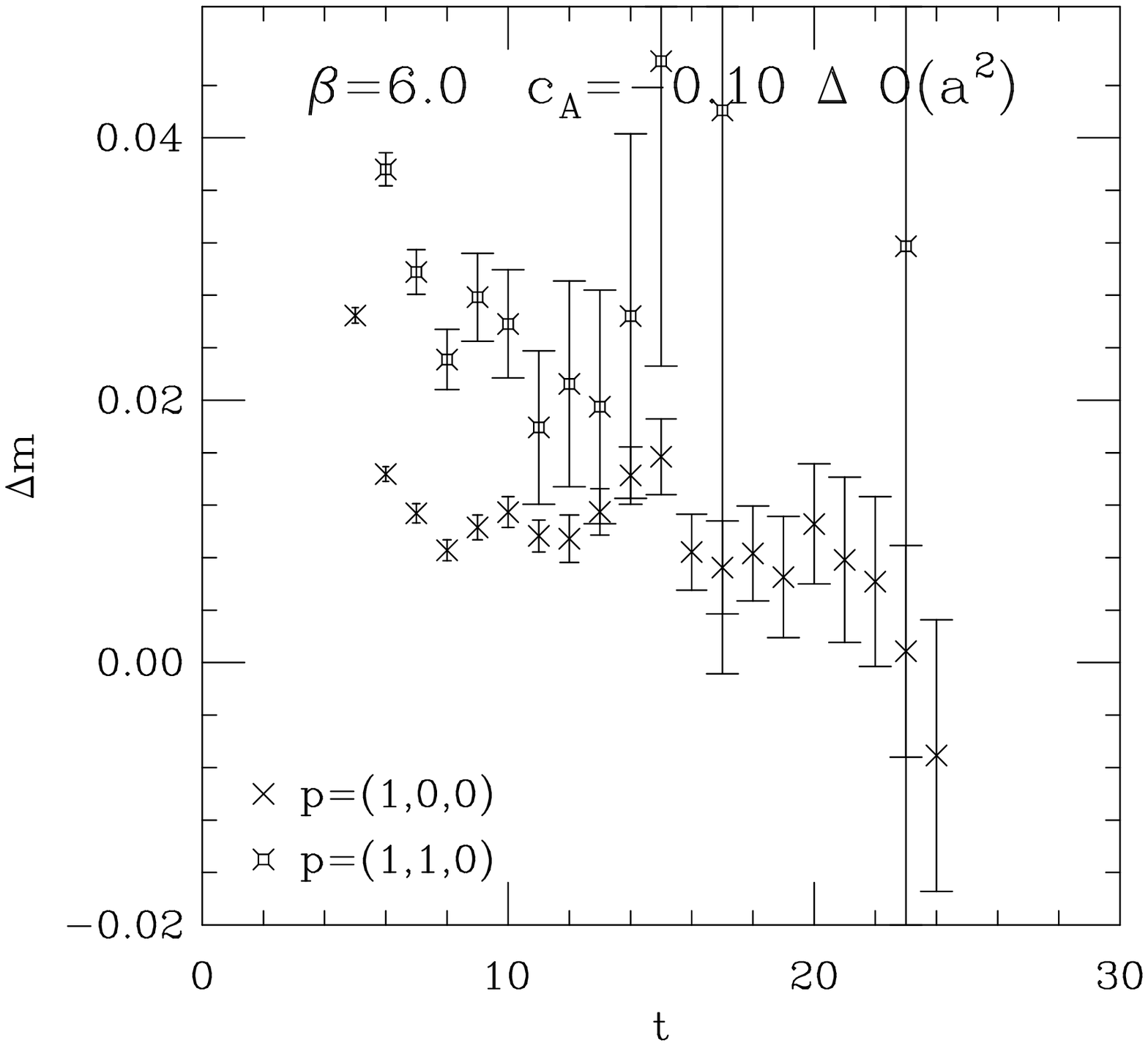,width=60mm}
\vskip-1cm
\caption{The difference in the bare quark mass determined at zero and
finite momentum, where standard lattice derivatives are
used.\label{diffa}}
\end{center} 
\vskip-0.5cm
\end{figure}

\begin{figure}
\begin{center}
\epsfig{file=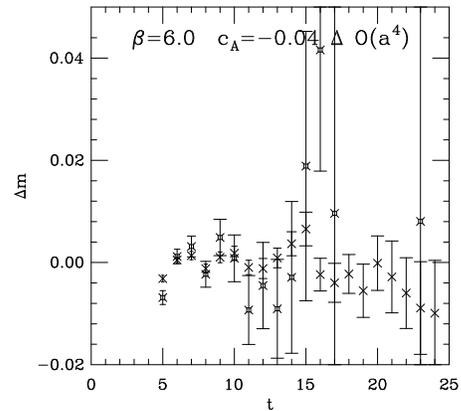,width=60mm}
\vskip-1cm
\caption{The difference in the bare quark mass determined at zero and
finite momentum, where improved lattice derivatives are
used.\label{diffb}}
\end{center} 
\vskip-0.5cm
\end{figure}

To see how well the discretisation effects are reduced we plug these
values back into $r(t)+c_As(t)$ for LL correlators and compare with
the unimproved case. At $\beta=6.0$ we find that the plateau in $2m$
can be extended from $t=11$ to $t=6$ when $c_A=-0.10$ and $O(a^2)$
lattice derivatives are used. However, as shown in figure~\ref{cmass},
if one improves $\partial_\mu$, the discretisation effects can be
removed by the same amount but lower values of $c_A$ are
required. Clearly the value of $c_A$ and the form of the derivatives
significantly affects the value of $m$ extracted.  Furthermore, we
find that consistent values of $m$ at zero and finite momentum can
only be obtained if $O(a^4)$ or $O(a^6)$ derivatives are
used. Figures~\ref{diffa} and~\ref{diffb} show the difference in the
bare quark mass at zero and finite momentum for standard and improved
derivatives.

\section{Results for $C_A$}
\begin{figure}[t]
\begin{center}
\epsfig{file=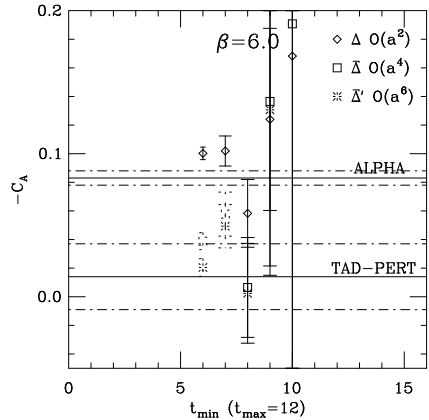,width=60mm}
\vskip-1cm
\caption{The results for $c_A$ at $\beta=6.0$ obtain using correlated fits with
$Q>.10$~(the dashed points have $0.10>Q>.01$). The data has been
averaged over positive and negative~(i.e. $t>25$) timeslices. The
errors are bootstrapped over 100 bootstrap samples. The result of the
ALPHA collaboration is shown as a horizontal line, where the dashed
lines indicate the numerical error. The tadpole-improved $1$-loop
perturbative result is also indicated, the error is taken to be
$1\alpha_p^2(\pi/a)$.\label{caresa}}
\end{center}
\vskip-0.5cm
\end{figure}

\begin{figure}[t]
\begin{center}
\epsfig{file=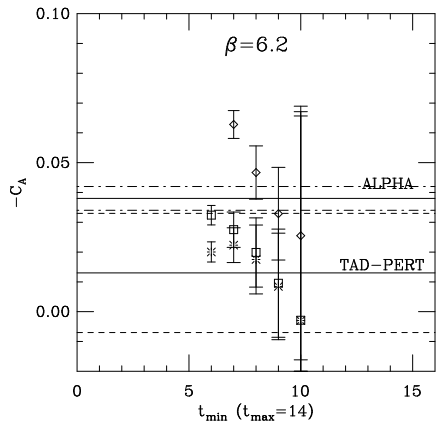,width=60mm}
\vskip-1cm
\caption{The results for $c_A$ at $\beta=6.2$. The figure is labelled in the same way as figure~\protect\ref{caresa}\label{caresb}}
\end{center}
\vskip-0.5cm
\end{figure}

Results are shown in figures~\ref{caresa} and~\ref{caresb}.
The main points are:
\begin{itemize}
\item A reliable estimate for $c_A$ is indicated by
a stable value with $t_{min}$. At $\beta=6.0$, the statistical errors
grow rapidly for $t_{min}>8$, although the results are reasonably
stable for $t_{min}=6-8$. The situation is similar at $\beta=6.2$.
\item Clearly $c_A$ decreases when the temporal derivatives are
improved. The decrease is less going from derivatives correct to
$O(a^4)$ to those correct to $O(a^6)$, as expected.
\item $c_A$ obtained using improved derivatives is below the
value obtained by the ALPHA collaboration. For the $O(a^6)$-correct
derivatives $c_A$ is close to the 1-loop tadpole-improved
value~(although the perturbation theory has not been performed with
improved derivatives). In principle, our results can differ with those
of reference~\cite{alpha} by $O(\Lambda_{QCD}a)\sim .300/2.1 =.14$ at $\beta=6.0$ and
$.10$ at $\beta=6.2$.
\item The results can be compared 
to those obtained by Bhattacharya et. al.~\cite{lanl}. They obtain
$c_A=-0.02(2)(2)$ at $\beta=6.0$ and $-0.033(4)(3)$ at $\beta=6.2$ for
$\kappa_c$ with a significantly smaller number of configurations. Our
results are at finite $\kappa$ around $\kappa_{strange}$. A
preliminary study indicates that the light quark mass dependence of
$c_A$ is small.
\end{itemize}

\section{Conclusions}
We have applied the method of reference~\cite{lanl} to extract the $O(a)$
improvement coefficient, $c_A$. We find that the accuracy and
reliability of this method is limited by the small range of timeslices
over which a stable value of $c_A$ is found. Nevertheless, we have
shown that the value of $c_A$ obtained depends significantly on the
discretisation chosen for the lattice derivatives appearing in the
PCAC relation. $c_A$ reduces as the lattice derivatives are improved,
and lies between the value obtained by the ALPHA collaboration and the
1-loop tadpole-improved value. Furthermore, consistent values for the
bare quark mass at zero and finite momentum can only be obtained for
derivatives correct to $O(a^4)$ and above.

These results suggest a similar study should be performed within the
Schr\"odinger functional approach. In the future, we plan to
investigate the quark mass dependence of $c_A$ and check the scaling
properties of the renormalised light quark mass and $f_\pi$ obtained
using our values of $c_A$. 
   
\section{Acknowledgements}
We thank T.Bhattacharya and R.Gupta for discussions on the details of
their method. S.~Collins has been supported by a Royal Society of
Edinburgh fellowship.

\end{document}